\documentstyle[twoside,fleqn,espcrc2,psfig]{article}

\newcommand{\real}{{\sf I}\kern-.12em{\sf R}}

\title{Resummation of Cactus Diagrams in Lattice QCD, to all Orders}

\author{H. Panagopoulos\address{Department of Natural Sciences, University of Cyprus,
        P.O.Box 20537, Nicosia CY-1678, Cyprus.}, 
        E. Vicari\address{Dipartimento di Fisica dell'Universit\`a and I.N.F.N., 
        via Buonarroti 2, I-56127 Pisa, Italy. }}

\begin{document}

\begin{abstract}
We show how to perform a resummation, to all orders in perturbation theory,
of a certain class of gauge invariant tadpole-like diagrams in Lattice
QCD. These diagrams
are often largely responsible for lattice artifacts. 
Our resummation leads to 
an improved perturbative expansion. Applied to a number of cases of interest,
e.g. the lattice renormalization of some two-fermion operators, this
expansion yields results remarkably close to corresponding 
nonperturbative estimates. 
We consider in our study both the Wilson and the clover action for
fermions.

\end{abstract}

\maketitle

\section{INTRODUCTION}

Ever since the earliest days of lattice field theory, one problem
present in most numerical simulations has been the calculation of
corrections induced by renormalization on Monte Carlo
results. Several methods have been used to address it.
Perturbation theory provides in principle a methodical
means of calculating renormalization
functions, operator mixing coefficients, etc. Its drawbacks lie in its
asymptotic nature, and in that it is a formidable task on the lattice,
which places severe limitations on the order to which it can be
carried out; indeed, at present, exact calculations in perturbative
lattice QCD reach only two loops (for 2-point
diagrams)~\cite{lw234,afp,cfpv}  and three loops (for vacuum
diagrams)~\cite{afp2}. 
In recent years there have been considerable efforts in refining the perturbative 
computations introducing recipes motivated by mean field and tadpole resummation
arguments~\cite{lepage}.   
Various nonperturbative, numerical approaches to renormalization functions
have also been devised and there has been recent progress both in
their range of applicability and in their precision~\cite{romani,M-P-S-T-V,luescher}. 
Nonperturbative methods, such as those of Refs.~\cite{romani,M-P-S-T-V,luescher}
are in general preferable to approximations based on perturbative calculations, due
to their better controlled systematic errors
($O(a)$ against $O(g_0^n)$). However, improved perturbative estimates
are still quite useful. They indeed provide important consistency checks.
Further, in those cases where nonperturbative methods
are difficult to implement, perturbative methods remain
the only source of quantitative information.

\medskip
\centerline{\psfig{figure=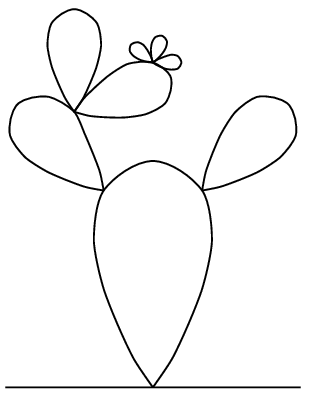,width=3.5truecm}}
\centerline{{\bf Figure 1: A cactus}}

\medskip

We present an improvement of lattice perturbation
theory, which results from a resummation to all orders of a certain
class of diagrams, dubbed ``cactus'' diagrams~\cite{cactus1,cactus2}. 
Briefly stated, these
are tadpole diagrams which become disconnected if any one of their
vertices is removed (see Figure 1). Our original motivation was the well known
observation of ``tadpole dominance'' in lattice perturbation theory.
This observation must clearly be taken with
a grain of salt: 
One-sided inclusion of tadpoles can ruin desirable
partial cancellations between tadpole and non-tadpole diagrams; worse,
their contribution is gauge dependent. 
The class of terms we propose
to resum circumvents the latter objection since, as we shall see, it
is gauge invariant; it also overcomes the former objection in known
cases. Given its diagrammatic nature, our
method leads to results which can be systematically improved, order by
order. 
Applied to a number of cases of interest, the cactus resummation yields 
remarkable improvements when compared with the available nonperturbative estimates.
As regards numerical comparison with other improvement schemes, such as
boosted perturbation theory~\cite{lepage}, our method fares equally well on all the
cases studied.

\section{\bf THE CALCULATION}

Consider the standard Wilson action for $SU(N)$ lattice gauge fields:
$$
S = {1\over g_0^2} \sum_{x,\mu\nu} {\rm Re}\, {\rm tr} \left(1-U_{x,\mu\nu}^\Box
\right)
$$
$U_{x,\mu\nu}^\Box$ is the usual product of link variables around a
plaquette in the $\mu{-}\nu$ plane with origin at $x\,$; in standard
notation:
$$
U_{x,\mu\nu}^\Box = e^{i g_0 A_{x,\mu}} e^{i g_0 A_{x+\mu,\nu}} e^{-i
g_0 A_{x+\nu,\mu}} e^{-i g_0 A_{x,\nu}},
$$
where $A_{x,\mu} = A_{x,\mu}^a T^a$.
By the Baker-Campbell-Hausdorff (BCH) formula we have:
\begin{eqnarray}
U_{x,\mu\nu}^\Box &=& \exp\left\{i g_0 F_{x,\mu\nu}^{(1)} + {\cal O}(g_0^2) \right\} 
\nonumber \\
F_{x,\mu\nu}^{(1)}&\equiv&A_{x,\mu} + A_{x+\mu,\nu} -
A_{x+\nu,\mu} - A_{x,\nu} \nonumber
\end{eqnarray}

The diagrams which we propose to resum to all orders will be cactus
diagrams made of vertices containing $F_{x,\mu\nu}^{(1)}\,$. Let us
see how such diagrams will dress the gluon propagator; we write:
$$
\psfig{figure=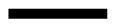,width=1truecm}
= \psfig{figure=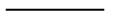,width=1truecm} +
\psfig{figure=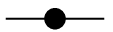,width=1truecm} +
\psfig{figure=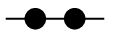,width=1truecm} + \cdots
$$
where the one-particle irreducible piece is given by the recursive
equation:
$$
\psfig{figure=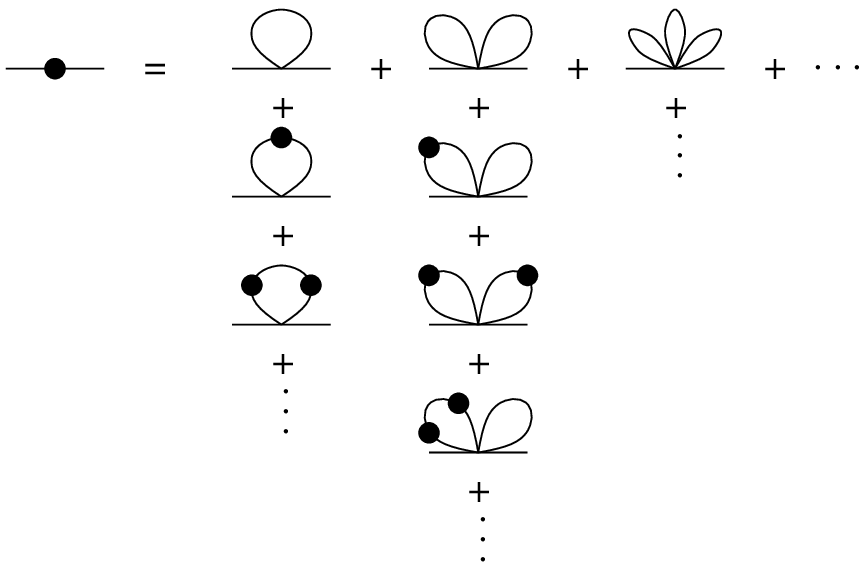,width=7truecm}
$$
The presence of only
$F_{x,\mu\nu}^{(1)}$ in the vertices ensures that the longitudinal parts of all
propagators cancel, so that the effect of dressing is the same in
all covariant gauges. 
Thus, we can write:
$$
\psfig{figure=propirr1.eps,width=1truecm} = w(g_0) \cdot 
\psfig{figure=propbare.eps,width=1truecm}
$$
Indeed, the dressed propagator will become a multiple of the bare
transverse one, where the factor $w(g_0)$ will depend on $g_0$ and
$N$, but not on the momentum. 
The above diagrammatic relations can be turned into an algebraic
equation for $w(g_0)$ that can be easily solved numerically.

Similar to the dressing of the propagator, one dresses 
the action vertices, operator insertions, etc.
Moreover these calculations can be extended~\cite{cactus2}  
to the case of the clover improved
action formulation of lattice QCD~\cite{clover}, 
which is widely used in numerical
simulations in order to reduce scaling corrections.

Cactus resummation may be applied either to bare quantities or to
quantities which have been calculated to a given order in perturbation
theory; thus contributions which are not included in the resummation
can be reintroduced in a systematic manner.

\section{\bf APPLICATIONS OF CACTUS RESUMMATION}

The resummation of the cactus diagrams
can be applied to the calculation of
the renormalization of some lattice operators.
Approximate expressions of the lattice renormalizations are
obtained by dressing the corresponding one-loop
calculations. 

As first example~\cite{cactus1}
we consider the renormalization
$Z(g_0)$ of the topological charge density lattice operator
$$
Q(x) =
-{1\over 2^4\times 32\pi^2}
 \sum_{\mu,\nu,\rho,\sigma=\pm 1}^{\pm 4}
\varepsilon^{\mu\nu\rho\sigma} {\rm tr} \{U_{x,\mu\nu}
U_{x,\rho\sigma} \}
$$
using the Wilson action and for $SU(3)$.
$Z(g_0)$ is a finite function of $g_0$,
going to one in the limit $g_0\to 0$, and that is much
smaller than one in the region $g_0\simeq 1$, where 
Monte Carlo simulations
using the Wilson action are actually performed.
In perturbation theory one has~\cite{C-D-P}
$Z(g_0) = 1-0.908 g_0^2+O(g_0^4)$,
leading to the estimate $Z(g_0=1)\simeq 0.092$.
The large one-loop coefficient suggests already that
perturbation theory can hardly provide  an acceptable
estimate of $Z(g_0)$ for $g_0\simeq 1$ without some kind of resummation.
Indeed the heating-method~\cite{D-V},
which does not rely on perturbation theory,
gives the estimate 
$Z(g_0=1)=0.19(1)$.
Cactus dressing of the  one-loop expression leads to
$$
Z_(g_0^2)\approx \left[ 1 {-} w(g_0)\right] \times  
\left[ 1 {-} w(g_0) - 0.491 g_0^2\right]
$$
Since $w(g_0=1)\simeq 0.250$, one finds
$Z(g_0=1)\simeq 0.193$, in agreement with the heating-method result.
A similar estimate is obtained 
by ``mean-field'' improving~\cite{lepage} the one-loop
calculation:
$Z \simeq P^2\left( 1 - 0.241 g_P^2\right)= 0.209$,
where $P\simeq 0.5937$ is the plaquette and 
$g_P^2=g_0^2/P$ a new coupling. 
Further confirmation~\cite{cactus1} 
of the validity of the cactus resummation 
has been achieved for $SU(2)$.

One can also apply cactus resummation to the lattice renormalization
of fermionic operators. 
In particular we consider 
the renormalizations $Z_V, Z_A$ of isovector fermionic currents
using the clover action~\cite{cactus2}.
Their one-loop result is known~\cite{Cap-et-al-1}, 
$Z_{V,A} = 1 + g_0^2 z_{V,A}(c_{\rm SW}) + \ldots$
Cactus dressing amounts to: 
$$1 + {g_0^2\over 1-w(g_0)}\, z_{V,A}\left((1-w(g_0)) c_{\rm SW}\right) + \ldots$$

\psfig{figure=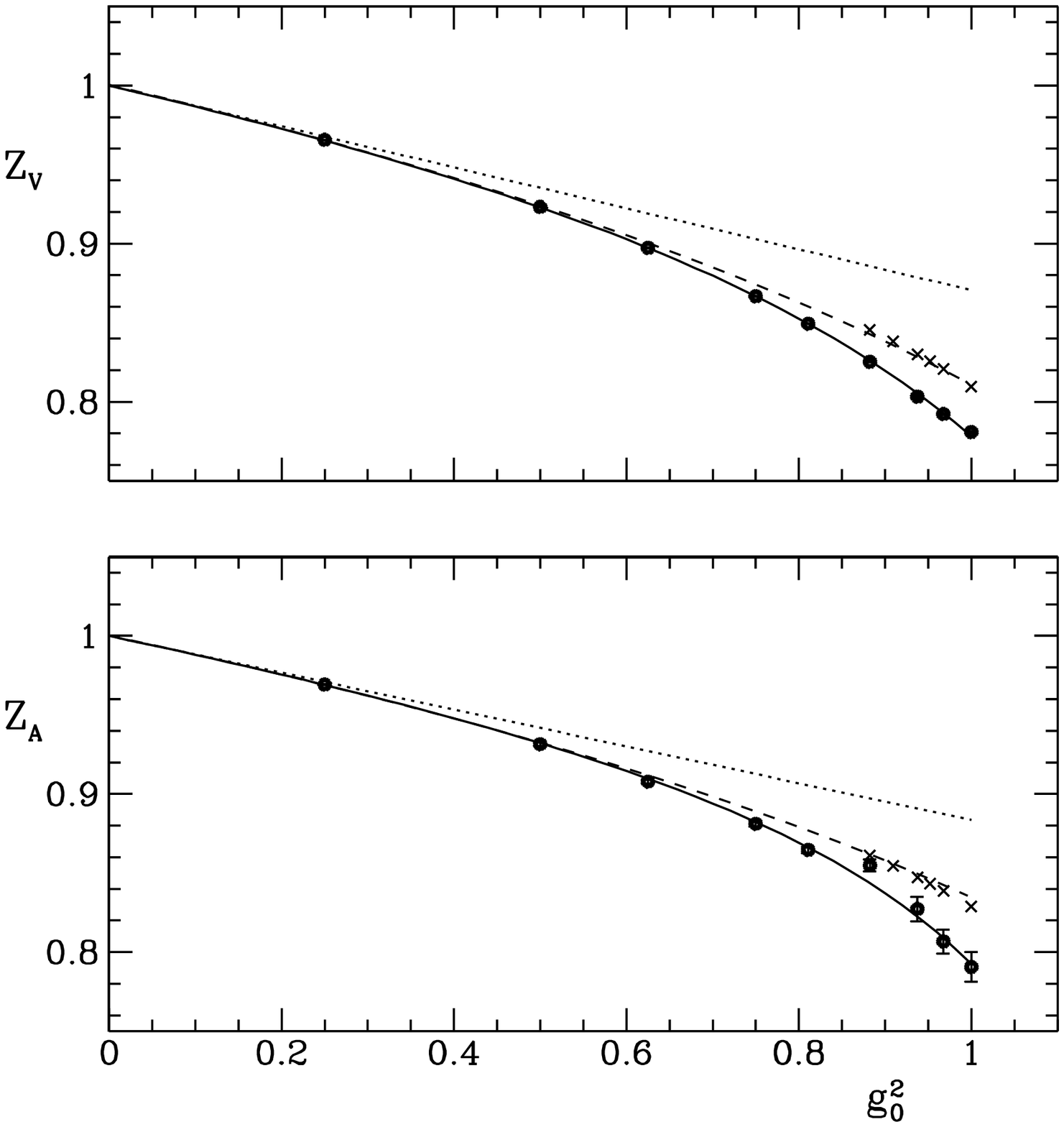,width=8truecm}

In the above figures we show results
for $Z_V$ and $Z_A$ (from Ref.~\cite{luescher}), coming from numerical simulations (filled
circles, fitted by a solid line), bare perturbation theory (dotted lines) and ``mean
field improved" perturbation theory~\cite{lepage} (crosses). The 
superimposed dashed lines are our results from cactus dressing.
As this comparison shows, the dressed one-loop expressions constitute a 
remarkable improvement with respect to the simple one-loop calculation.


\end{document}